\documentclass[11pt,twocolumn]{article}
\usepackage{amsmath}
\usepackage{amssymb}
\usepackage{authblk}
\usepackage{bm}
\usepackage{bbold}
\usepackage{hyperref}
\usepackage{subcaption}
\usepackage{graphicx}
\usepackage[style=apa,sorting=nyt,backend=biber,natbib]{biblatex}
\usepackage{lscape}
\usepackage{afterpage}
\usepackage{xcolor}
\newcommand{\pdiff}[2]{\frac{\partial#1}{\partial#2}}

\newcommand{\vF}[0]{\mathbf{F}}
\newcommand{\avep}[2]{\left<#1\right>_{#2}}

\begin{document}
\twocolumn[
\begin{@twocolumnfalse}

\title{Evolutionary path dependence of semantic complexity}
\author{Elliot M. Butterworth$^{1}$, Tim Rogers$^{1}$, Matthew A. Wills$^{2}$}
\affil{$^{1}$ Department of Mathematical Sciences, University of Bath, Bath, BA2 7AY}
\affil{$^{2}$ Centre for Evolution, University of Bath, Bath, BA2 7AY}
\date{}
\maketitle

\begin{abstract}
    Attempts to quantify biological complexity often consider intrinsic structural properties at a chosen hierarchical level and resolution, such as counts of body parts and their degree of differentiation.  These measures are inherently \emph{syntactic}, being concerned with the information needed to specify an arrangement rather than the biological functions performed.  Syntactic complexity alone is therefore not sophisticated enough of a measure to fully address the role of complexity as either a driver or consequence of evolution.  We propose to study the counterpart, \emph{semantic} complexity: the subset of structural features whose variation has a measurable effect on organismal fitness.  We illustrate this distinction in a simple mathematical model of tagmosis with functional constraints, symmetry breaking, and specialisation.  We find that the total syntactic complexity evolved as selection drives lineages toward globally optimal fitness is path-dependent, revealing two evolutionary modes: a driven mode, in which semantic and syntactic complexity rise together, and an entropic mode, in which syntactic complexity drifts upward under a near-neutral evolution.  Historical contingencies in early specialisation, combined with multi-optima fitness landscapes, govern how long lineages stay in each mode.  Those on paths that do not lead directly to the highest-fitness states remain in the driven mode for longer and can eventually reach comparable fitness, but only by evolving morphologies with greater syntactic complexity.
\end{abstract}
\end{@twocolumnfalse}
\vspace{1cm}
]

\section{Introduction}
One of the most striking features of the history of life is the apparent increase in complexity across a wide range of biological systems.  Over evolutionary time, simple molecules gave rise to metabolic networks \citep{Wagner2001}, unicellular organisms to multicellular ones \citep{Grosberg2007}, undifferentiated tissues to specialised organs \citep{Carroll2005}, and solitary individuals to integrated colonies and ecosystems \citep{Szathmáry1995,Bourke2011}.  These trends are neither universal nor unidirectional, but the emergence of more structurally and functionally elaborate systems is a recurrent theme across many lineages and levels of organisation \citep{McShea1996,Knoll2000,Bonner2004}.

Ironically, it is precisely this enormous organisational disparity of forms that makes it nearly impossible to quantify complexity in a universal framework, thereby hampering comparative analyses \citep{Duclos2019,Rebout2021}.  Consequently, biological complexity has been defined in many ways, reflecting both the hierarchical organisation of living systems and the different properties that can vary within each level, each expressing an intrinsic property of an organism.  Definitions variously encapsulate the number and diversity of components, their degree of differentiation, the nature of their interactions, the presence of nested organisation, modularity and network architecture \citep{McShea1996,Bonner2004,Carroll2005,Duclos2019}.  These properties need not vary together.  Complexity may differ among hierarchical levels and may even be multidimensional at a specified level: the number of parts, their differentiation and the regularity of their arrangement, for example, can vary partly independently \citep{McSheaBrandon2010,Rock2025}.

Despite the diversity of definitions, these complexity measures share an important feature: they describe the intrinsic structure of an organism.  Similarly, mathematical concepts such as Kolmogorov complexity \citep{Kolmogorov1965} provide, for a specified representation, an invariant (context independent) measure of the information required to describe an object.  Using terminology analogous to that of \citep{Kolchinsky2018}, we refer to these intrinsic, structure-based properties as \emph{syntactic} complexity.  This term does not resolve the longstanding problem of how complexity should best be quantified.  Rather, it distinguishes that problem from a second, equally pressing problem: structural complexity can be measured without establishing whether, or to what extent, it contributes to survival and replication.

To address this problem, we will distinguish total structural complexity (syntactic complexity) from that component of complexity whose variation makes a measurable difference to the fitness of an organism \citep{Veit2025}.  Again, following \citet{Kolchinsky2018}, we refer to the latter as \emph{semantic} complexity.  Whereas syntactic complexity describes the organisation and features of the organism itself, semantic complexity is necessarily contextual: it depends upon what those features in that organisation enable the organism to do in a particular environment.  Selection may attune organisms to those environments, but its outcomes are conditioned by the variation available, developmental constraints, genetic drift and the accumulated contingencies of lineage history.  Whether a particular component of syntactic complexity is also semantic therefore depends on both intrinsic and extrinsic contexts.  Importantly, fitness need not increase monotonically with syntactic complexity.  Organisms with insufficient complexity may be unable to meet particular functional demands or exploit opportunities in their environment, whereas complexity beyond that required may provide no further advantage and may instead merely impose developmental, energetic or other costs \citep{Wang2010,OMalley2016}.  The same amount or form of syntactic complexity may consequently have different semantic significance in different organisms and environments.

Many organisms evolve specialised morphological features that perform particular functions.  Importantly, the entire body does not need to contribute to such functions: it is sufficient for a specific unit or region to undertake it.  The functional capability of the organism then depends on how effectively that specialised unit fulfils its role.  In many arthropods, only a single pair of appendages is specialised to perform a particular function, while the rest of the limbs serve unrelated roles.  Mantis shrimps (Stomatopoda) provide a striking example.  Their raptorial appendages have evolved into highly modified clubs or spears capable of delivering one of the fastest and most forceful blows in the animal kingdom, used exclusively for prey capture and defence, while the other thoracic limbs are specialised for walking or handling food \citep{Patek2004}.  A comparable pattern is seen in praying mantids (Mantodea), where the prothoracic limbs are adapted as raptorial appendages equipped with spines and grasping structures to seize prey, while the meso- and meta-thoracic limbs function only in locomotion \citep{Prete1990}.  Across crustaceans, one or more anterior thoracic appendage pairs have repeatedly been transformed from walking legs into feeding appendages, associated with changes in the boundaries of Hox-gene expression \citep{Averof1997}.  More posterior thoracic limbs retain locomotor functions.  This demonstrates how a developmental change can release particular members of a serially homologous series from their ancestral identity, permitting a restricted subset to specialise while the rest remain comparatively similar.  In contrast, many functions, including locomotion, depend not on the specialisation of a single appendage but on coordinated activity distributed across multiple appendages or body regions.  In many crustaceans, effective swimming arises from the metachronal beating of multiple abdominal pleopods, with propulsion emerging only from precise phase relationships among appendages \citep{Alben2010}.  Myriapods likewise generate travelling waves of leg movement, requiring coordination across dozens of limbs for efficient crawling or burrowing \citep{Diaz2023}.  Trilobites are thought to have employed similar metachronal patterns, with trackway evidence and reconstructions showing locomotion as the product of coordinated thoracic limb activity \citep{Esteve2023}.  Even in annelids, peristaltic crawling and burrowing depend on rhythmic alternation of contraction and anchoring across many segments, with no single segment sufficient to produce forward motion \citep{Darwin1881}.

Inspired by such observations and to explore the interplay of syntactic and semantic complexity with evolutionary dynamics, we investigate a mathematical model of tagmosis \citep{Fusco2013}, the evolutionary differentiation of repeated body segments into distinct functional regions, termed tagmata (see Figure~\ref{fig:Figure1}~(a)).  It is often associated with increased differentiation among the serially homologous somites that make up the body, together with their appendages.  Although segments may originate from a relatively uniform developmental ground plan, they can acquire distinct morphologies and functions over time, leading to regional specialisation.  This is a process in which symmetry is broken, releasing constraints that previously enforced identical adaptation of segments.  We consider this process as a paradigmatic example of a system where syntactic and semantic complexity can vary, and investigate a model of tagmosis and differentiation to explore the way in which complexity can evolve.  We quantify the extent to which repeated structures diverge in form and function, and examine how their division of labour relates to fitness and evolutionary optimality \citep{Brinkworth2023,McSheaBrandon2010}.  The grouping of body segments into functional units and the subsequent differentiation of those units has an intuitive impact on complexity \citep{Cisne1974,Wills1998,Adamowicz2008}, which maps to a precise mathematical framework that we specify below.

Our model incorporates two distinct kinds of constraint that together delimit the morphologies available to evolution \citep{Kempes2019}.  First, each functional unit (a subset of a morphology that can perform functions) has a finite functional capacity: improving its performance in one function reduces the available capacity for performing others, creating a trade-off between generalisation and specialisation.  This functional capacity constraint restricts each unit to a bounded region of morphospace.  Second, symmetry constraints require sets of serially homologous units to share the same morphology and functional capacities.  Such units cannot vary independently, reducing the dimensionality of the whole-organism morphospace and limiting the possible division of labour among them.  Tagmosis is represented as the progressive breaking of these symmetries.  Releasing a symmetry constraint enlarges the accessible morphospace, allowing formerly identical units to differentiate and specialise.  Some of the newly accessible morphologies may improve functional performance and fitness, whereas others may increase syntactic complexity without providing any further selective advantage. 

We find that the emergent evolutionary dynamics are characterised by two distinct modes under which complexity can increase: a driven mode where higher complexity is selectively advantageous and an entropic mode where greater complexity enlarges the optimal morphospace, but does not provide access to higher fitness.  Our results also show that there are multiple evolutionary paths leading to optimal fitness and, crucially, that the path taken determines how long a lineage spends evolving under each of the two modes, affecting the complexity of the organism first evolving an optimally fit morphology in a given lineage.  We discover that lineages evolving sub-optimal morphologies due to unfavourable initial specialisations must remain in the driven mode for longer, requiring the evolution of greater complexity in order to produce organisms with equal fitness to those whose lineages evolve favourable morphologies earlier in their evolutionary history.

\section{Results}
\subsection{Mathematical framework}
\begin{figure*}[htb!]
    \centering
    \includegraphics[width=\textwidth]{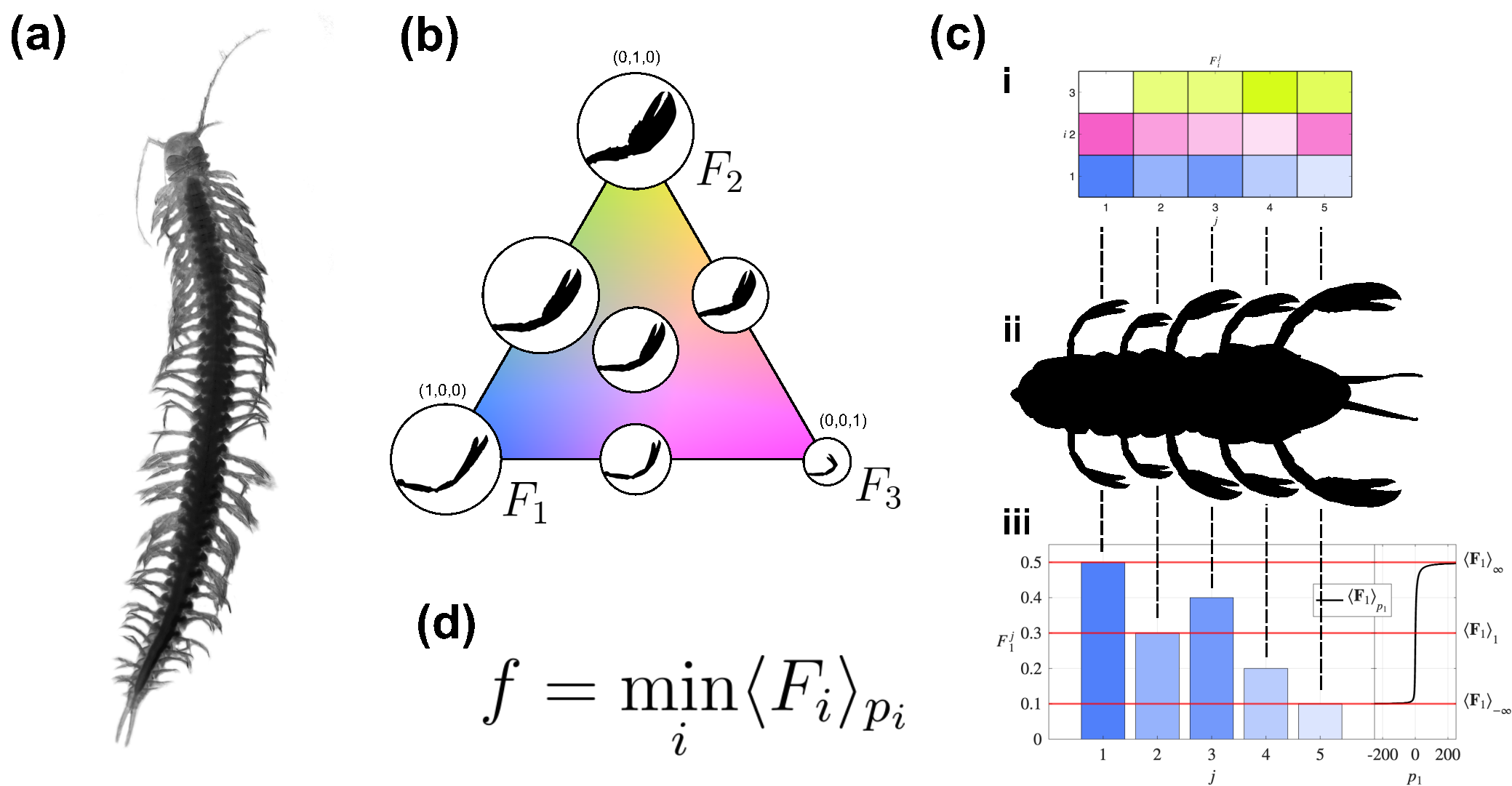}
    \caption{(a) Segmentation and differentiation of form and functional role of paired limbs in the remipede \citep[image adapted from Fig.~1;][]{Pinheiro-Junior2024}.  (b) A model simplex for single-limb morphology allowing specialisation towards one of three functions.  With $F_i^j$ denoting the ability of limb-pair $j$ to perform function $i$, a global constraint $\sum_iF_i^j\leq1$ implies a choice between specialisation (the corners of the simplex) or general purpose limbs (the centre).  (c)(i) The morphology of all limb-pairs is encoded in the matrix $F$ with rows corresponding to function and columns to body position (ii).  (iii) The overall capability of the organism to perform function $i$ is computed as the norm $\langle F_i\rangle=\left(\sum_j\left(F_i^j\right)^{p_i}\right)^{1/p_i}$; varying the exponent $p_i$ interpolates between min, max and additive contribution.  (d) Total organismal fitness is taken as the minimum of capability across all functions.}
    \label{fig:Figure1}
\end{figure*}

For the mathematical details of our model, see Section~\ref{sec:Methods_TheGeneralModel}.  Here, we outline the concepts underlying the model.  Our general model considers a hypothetical organism divided into functional units that collectively perform a set of functions required for survival and reproduction.  Since all functions are assumed to be essential, organismal fitness is limited by its poorest functional performance (see Figure~\ref{fig:Figure1}~(d)).  Some functions may require only specialisation of a single functional unit to be performed well, others may require capability and integration across many functional units.  Mathematically, we can encode this property through the use of a weighted vector norm-like function that interpolates between min, max and additive contributions (see Figure~\ref{fig:Figure1}~(c)(iii)).  Each functional unit is subject to a limiting maximal functional capacity, imposed by morphological constraints, which prevents it from achieving optimal specialisation for all functions simultaneously.  This limited functional capacity means that full specialisation to one function may come at the expense of the ability to perform any other functions, realised mathematically as a position in a morphological simplex (see Figure~\ref{fig:Figure1}~(b)).  Syntactic complexity can vary depending on the number of functional units as well as the number of constraints.  Semantic complexity describes what aspects of any complex morphology contributes to greater fitness.

Starting from the morphology of an organism, we assume that it is possible to assess fitness in a given environment by evaluating its capacity to perform required functions in that environment:
\begin{equation*}
    \mathrm{Morphology}\quad\rightarrow\quad\mathrm{Capability}\quad\rightarrow\quad\mathrm{Fitness}
\end{equation*}
The mathematical version of this two stage deterministic map is presented in Section~\ref{sec:Methods_TheGeneralModel}.  To model the division of an organism into functional units by the process of tagmosis, we track the ability of every functional unit of a model organism to perform every function.  The two stage deterministic map from this set of functional abilities to the fitness of the organism can then be applied.  The first stage assesses the capability of the organism to perform each function by taking into account the ability of each functional unit.  The second stage assesses which function is performed the least effectively, defining this as the fitness (the limiting factor on the survival of the organism).  The mapping in the first stage can be different for each function (see Figure~\ref{fig:Figure1}~(c)(iii)).  Some functions can be performed effectively by specialised units whereas others may require adaptation across many units for effective performance.  Finally, the mapping can also include the preference for a given function to be performed by certain functional units over others.  Object manipulation tasks, for example, may be performed better by functional units in front of the eyes that can be seen easily.

Morphological structures have finite capacities due to limitations imposed by physical, developmental and historical factors \citep{McGhee2015,Oyston2015}.  Adaptations of a functional unit that improve one function may compromise its performance in others.  Such trade-offs can arise because alternative functions place conflicting demands on geometry, joint mobility, musculature or control.  As a deliberately minimal representation of these constraints, we assume that each functional unit has a finite, normalised capacity that can be distributed among the functions it performs.  Increasing its adaptation to one function therefore reduces the capacity available for adaptation to others.  There is a ``cost'' associated with each function that describes the relative functional capacity budget used to evolve equal levels of specialisation for each function.  In Section~\ref{sec:Methods}, we investigate the model with idealised parameters such that complete specialisation to one function takes up the entire functional capacity budget, preventing any other function from being performed by that specialised functional unit.  In general, such specialisation will not cause the loss of every other capacity: more general parameters allow functional units to specialise whilst also maintaining a baseline capacity for performing various other functions.

\subsection{Direct and indirect paths to optimal fitness}\label{sec:ThePathToOptimalFitness}
\begin{figure*}[htb!]
    \centering
    \includegraphics[width=\textwidth]{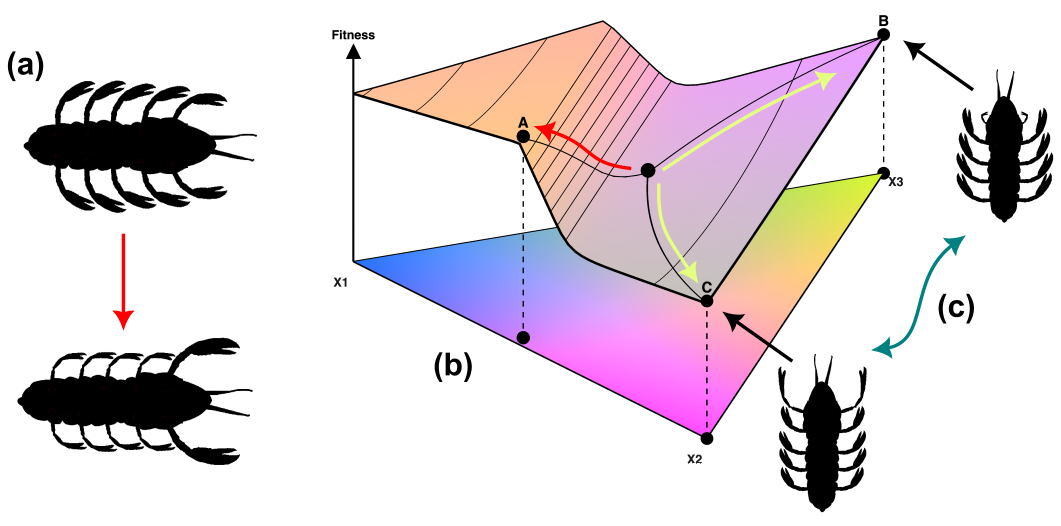}
    \caption{(a) The most simple organism with (locally) optimal fitness has all leg pairs performing generalist roles; the direct complexifying route to near global optimal fitness for these model parameters is differentiation of one leg pair to specialise in food gathering with the rear pairs all identical and balanced between locomotion and food handling.  (b) Illustration of the fitness landscape.  The base simplex denotes the morphology of the front leg pair, the height surface above shows the optimal fitness that could be achieved with that front leg morphology, and the colour of the surface matches the simplex location for the morphology of the rear legs.  The red arrow indicates the path uphill in fitness from central saddle (representing the simple symmetric organism) to the global optimum point A, illustrated in panel (a).  (c) Alternative paths uphill in fitness from the simple organism arrive at one of two local optima, B or C, at which the front legs are differently specialised.  These optima are not global and are connected by a path of near constant fitness along which the organism might evolve according to near neutral drift.}
    \label{fig:Figure2}
\end{figure*}

The direct path to optimal fitness is the one that requires the fewest complexifying mutations before an organism reaches the global optimum.  These mutations release constraints, providing access to more complex morphologies, some of which confer higher fitness than those available at a lower level of morphological complexity.  The number required to reach globally optimal fitness is dependent on the specific environment an organism lives in, defined by a fitness function.  The environment will have at least one maximum fitness (for a given body plan) which will be achievable with a single morphology or family of morphologies.  The point where globally optimal fitness is reached is also the point where further mutations would increase the syntactic complexity with no associated increase in the semantic complexity.  By following the direct path, organisms attain optimal fitness with the lowest possible syntactic complexity, which we define as the optimal semantic complexity.  Once at this point, further complexifying mutations can give the lineage access to a greater number of morphologies conferring optimal fitness, but will not give access to higher fitness.  Simplifying mutations on the other hand will be deleterious, since globally optimal fitness cannot be attained with complexity lower than the optimal semantic complexity.  Thus, evolution subsequent to first reaching globally optimal fitness will be (assuming no significant cost to complexity) a near-neutral exploration of the higher-complexity-biased space of optimal morphologies, with complexity increasing on average and the optimal semantic complexity acting as a lower bound.

Indirect paths to optimal fitness involve more than the minimum number of complexifying mutations, resulting in organisms within a lineage on such a path achieving optimal fitness with syntactic complexity higher than the optimal semantic complexity, that is, higher than the minimum syntactic complexity required to do so.  When a complexifying mutation occurs, releasing a constraint, the fitness landscape changes as previously linked functional units become able to evolve independent adaptations.  Figure~\ref{fig:Figure2} shows how the fitness optimum in the previous landscape becomes a saddle point in the new, higher dimensional landscape following such a complexifying mutation.  Selection can then lead the lineage towards either global or local optima in the new landscape.  Local optima arise in the model when a function can be performed effectively by a small number of specialised units.  After the first complexifying mutation, for example, the front limb pair is able to evolve independently of the rear four limb pairs that remain constrained to share the same morphological features.  If the independent front pair specialises in gathering (a function which can be performed well by a single functional unit, see Section~\ref{sec:Methods} for details), the remaining pairs can direct their functional capacity towards the other essential functions of locomotion and feeding.  This arrangement gives globally optimal fitness.  If instead the rear four pairs evolve to perform all of the gathering, it is the front pair of limbs that directs its functional capacity towards locomotion and feeding.  This arrangement represents an improvement over the optimal morphology for the lowest complexity organism whose limbs must all evolve identical generalist strategies.  It is, however, a lower fitness morphology than the one with a single front pair of limbs specialised to perform the gathering function, making it a locally optimal morphology only.  More generally, this represents the difference between concentrating a specialised function in a small subset of serial homologues and spreading the same specialisation redundantly across a larger set that remains developmentally coupled.

For a lineage which has evolved to a local fitness optimum, there can be a large fitness valley to cross to access the global fitness optimum, largely prohibiting this evolutionary trajectory.  A further complexifying mutation can, however, induce the same change to the fitness landscape as before, causing the local fitness optimum to become a saddle point.  The lineage can then be driven by selection away from this point and towards a higher fitness state.  Once again, this may be to a local fitness optimum (where this evolutionary scenario can play out again) or to the global fitness optimum.

At some point, even a lineage that evolves to the local fitness optimum following every complexifying mutation will reach the global fitness optimum.  The fitness landscape corresponding to a body plan's maximally complex morphology, for example, has no local fitness optima since no functional units are constrained to evolve identical adaptations.  Importantly, every indirect path to globally optimal fitness involves more complexifying mutations than the direct path such that any lineage following an indirect evolutionary path results in optimally fit organisms with syntactic complexity higher than the optimal semantic complexity.

\subsection{Modes of complexity evolution}
\begin{figure*}[htb!]
    \centering
    \includegraphics[width=\textwidth]{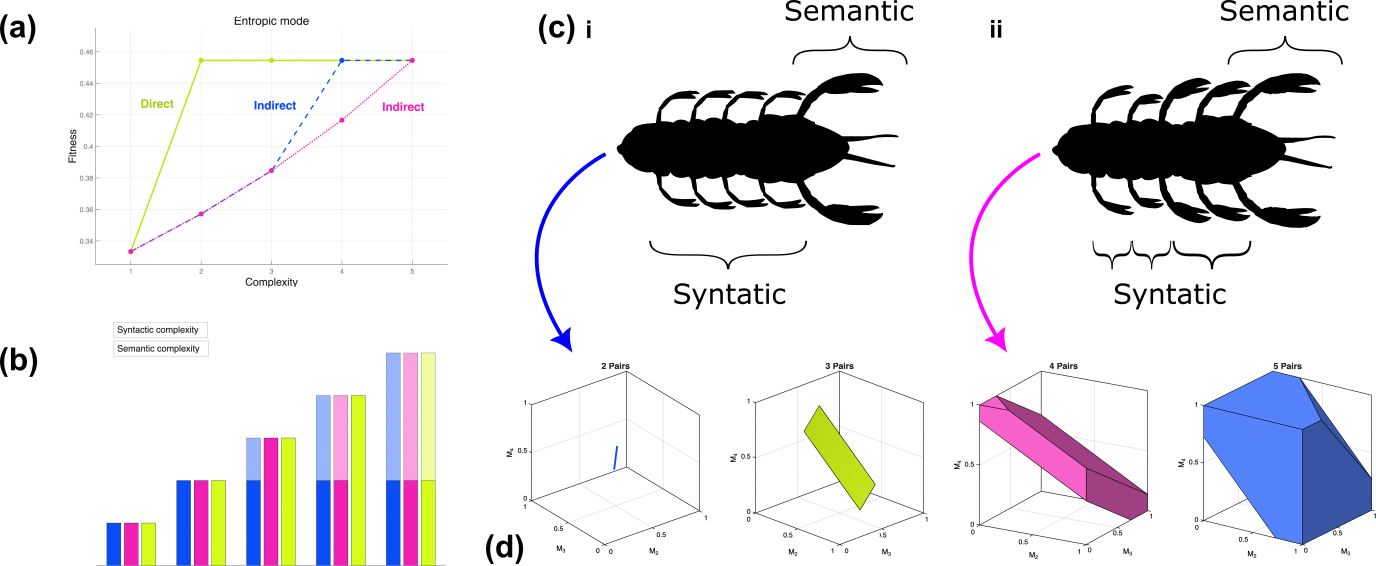}
    \caption{(a) Three evolutionary paths through stable fitness optima towards globally optimal fitness.  One is direct (solid green), the others (dotted pink and dashed blue) are indirect.  (b) The bar chart shows how syntactic and semantic complexity vary along each path.  As shown, the driven mode is characterised by complexity increases that provide access to higher fitness states, the entropic mode by complexity increases that only unlock (near) equal fitness states.  (c) Two equally fit organisms with equal semantic complexity (the innovation of differentiated large front claws).  Differentiation of the rear four leg pairs in (ii) increases the syntactic complexity of the organism without providing a strong selective advantage.  (d) The dimensionality of the optimal morphospace grows with each complexifying mutation, resulting in entropic evolution between various morphologies of near equal fitness.}
    \label{fig:Figure3}
\end{figure*}

Our model reveals that there are two distinct modes in which complexity can evolve.  The first is a driven mode: a complexifying mutation alters the fitness landscape in which a lineage is evolving, giving it access to higher fitness than was available before.  The second is an entropic mode: a complexifying mutation may alter the landscape in which a lineage evolves, but it does not provide access to higher fitness, instead, providing access to additional morphologies which have nearly equal fitness (see Figure~\ref{fig:Figure3}~(c)).  Thus, the lineage evolves within a near-neutral landscape of equally fit morphologies (at the resolution of selection), dominated by those of higher complexity, meaning random exploration leads to greater complexity on average.  The driven mode is characterised by increasing syntactic and semantic complexity.  Since complexifying mutations provide a lineage access to higher fitness, the complexity is semantic, as it unlocks selectively advantageous functional adaptation.  In the entropic mode however, complexifying mutations do not provide access to higher fitness: any increases in complexity are purely syntactic.  The transition between the two modes is the point at which complexifying mutations no longer increase semantic complexity, altering syntactic complexity only.

The evolutionary path taken by a lineage determines the time it spends in each of the two evolutionary modes.  By definition, the direct path leads a lineage out of the driven mode with the fewest complexifying mutations.  Indirect paths therefore cause a lineage to remain in the driven mode for more rounds of selection, which drives the evolution of higher syntactic complexity.  Assuming that the driven mode produces higher complexity in fewer selective sweeps than the entropic mode due to the additional evolutionary pressure, these results suggest that lineages taking an indirect path towards optimal fitness (evolving via local fitness optima) will produce higher complexity organisms earlier in their evolutionary history than lineages on the direct path.

\section{Discussion}
Measures of syntactic complexity, such as Kolmogorov complexity \citep{Kolmogorov1965}, quantify the minimum amount of information needed to describe an object independently of its context, and in many studies this provides a sufficient description.  Although this measure has proved valuable in many applications, we argue that it is insufficient in a biological setting.  In physics, for example, highly complex objects can arise from natural laws acting under specific conditions.  Such objects may represent the extreme tail of a distribution of possibilities that is itself governed by the underlying physical laws, while the frequency with which they are produced is determined by the likelihood of the conditions that give rise to them.  Detecting such an object can therefore provide information about the processes at work, or about the conditions in the dynamical system that is usually the focus of study \citep{Sharma2023}.  The complexity of the object is of interest only insofar as it reveals the conditions that must have existed to produce it.

Biological systems add a further layer to the problem of complex objects, because the process of replication induces additional selection \citep{Carroll2001}.  Physical and developmental processes delimit the range of complex morphological features that can arise.  Biology then focusses on the subset of those features which arise in a lineage and persist across generations, implying they are part of a morphology with high enough fitness to make it through the filter of selection.  However, persistence through generations does not imply that every component of a morphology is directly advantageous.  A feature may persist because it contributes to fitness, because it is neutral or only weakly deleterious, or because it cannot readily be altered independently of the rest of the organism.  Pleiotropy may cause changes to the feature also to affect other, advantageous traits, while developmental entrenchment may make later-developing structures dependent upon it.  In the latter case, a feature that was initially neutral (or advantageous only in an earlier context) may become costly to change or remove because subsequent evolution has built around it.  Selection then preserves the integrity of the larger developmental system, rather than necessarily favouring the focal feature for its present function.  Persistent complexity therefore raises a causal rather than merely descriptive question.  Does the focal structure itself improve fitness, is it retained as a correlated consequence of selection acting elsewhere, or has it become necessary through accumulated developmental dependencies?  These alternatives cannot be distinguished with syntactic complexity measures alone.

Complexity in biology cannot be understood independently of its role in survival and reproduction.  Measuring complexity in isolation captures only structure, which we call \textit{syntactic} complexity, and misses its functional significance.  \textit{Semantic} complexity addresses this gap, referring to the subset of structural features of an organism that have been shaped by natural selection because they contribute to survival and reproduction.  This distinction between absolute and functional complexity provides a useful framework for studying how and why complexity evolves, since only functionally beneficial complexity is expected to be maintained or increased by selection.  Complexity that offers no selective advantage may nonetheless persist if it is not strongly deleterious, often through neutral processes such as genetic drift.  Conversely, the loss of complexity along a lineage may reflect either direct selection against costly features or the gradual loss of traits that were not essential for effective survival and reproduction.

Our framework shows that complexity can evolve in two distinct modes.  In the driven mode, increases in complexity allow a lineage to evolve fitter morphologies.  In the entropic mode, increases in complexity give no access to greater fitness but instead expand the space of equally fit morphologies.  This leads to near-neutral evolution (assuming no significant cost to complexity) within a set of morphologies dominated by those of higher complexity, so that complexity increases on average.  In this entropic mode the optimal semantic complexity acts as a lower bound on the complexity that can be explored neutrally, since no morphology is less complex than this and also globally optimal in fitness.  Differences in the cost to a functional unit of performing each function can also be incorporated into the model, with specialisation towards more costly functions limiting adaptation to other functions more than specialisation towards less costly ones.

Framing tagmosis as the release of symmetry constraints connects our model to the zero-force evolutionary law (ZFEL).  The ZFEL identifies a general tendency for parts that vary with some independence to become increasingly differentiated, unless that tendency is opposed by selection or constraint \citep{McShea2005,McSheaBrandon2010,Brandon2010,McShea2019}.  In our model, symmetry constraints suppress this tendency by requiring several functional units (limb pairs) to share the same morphology.  Releasing such a constraint allows previously coupled units to vary independently, increasing the dimensionality of the accessible morphospace and creating the possibility of greater syntactic complexity.  The ZFEL does not explain why the coupling is released; rather, it describes the tendency towards differentiation once that independence has been established.  The consequences of this new freedom depend on the fitness landscape.  Where constraint release provides access to higher-fitness morphologies, selection directs the newly independent units towards different functional roles.  Differentiation then contributes to division of labour, and syntactic and semantic complexity increase together in the driven mode.  Where globally optimal fitness has already been reached, further constraint release instead enlarges the morphospace containing equally fit morphologies.  Nearly neutral exploration of this expanded space can then increase differentiation without improving function, corresponding to the entropic mode.  This second case is the closest counterpart of the ZFEL expectation: syntactic complexity increases because independently varying parts tend to diverge, not because differentiation itself is advantageous.  The present model abstracts from these dependencies by allowing symmetry constraints to be released without collateral effects on other aspects of development.  In biological systems, pleiotropy and developmental entrenchment may make such release costly or difficult, and may also prevent previously acquired complexity from being lost.  They therefore provide an additional source of path dependence: the fitness consequences of a structural change depend not only on its present function, but also on the genetic and developmental dependencies accumulated earlier in the lineage.

The driven mode produces high-complexity states in fewer rounds of selection than the entropic mode, because a population of higher-complexity mutants is strongly selected for, whereas in the entropic mode the complexifying mutations are neutral.  Evolving along an indirect path therefore means evolving in the driven mode for more generations, which produces higher-complexity organisms earlier in a lineage than when the optimal fitness state is reached in fewer rounds of selection and the lineage passes quickly into the entropic mode (Figure~\ref{fig:Figure3}~(a)).  Comparing a lineage that evolved along the direct path with one that evolved along an indirect path would show an optimal adaptation appearing earlier in the former, where it would then be retained in all subsequent species despite variation in other aspects of the morphology.  In the latter, the same adaptation would be expected to appear later, and in several species spanning a range of morphotypes that acquire it later along an indirect path.  Thus, there are two signatures of sub-optimal evolutionary branching: late acquisition of an optimal adaptation that may be seen early in the evolutionary history of other lineages, and higher complexity organisms relative to other sets of related organisms from lineages of similar age.

To end, we note that despite overall trends of complexification, greater complexity is not always advantageous.  Most organisms remain relatively simple \citep{Bar-on2018,deVargas2015,Hoshino2020,Mora2011}, and there are numerous examples in which simplification appears to be selected for \citep{Auerbach2014,Soyer2006,Veit2025}.  As with many macroevolutionary and macroecological generalisations or ``rules'' (for example Cope's Rule \citep{Solow2010}, Dollo's Law \citep{Dollo1893}, Williston's Law \citep{Williston1914}, Bergmann's Rule \citep{Ashton2000} and Rapoport's Rule \citep{Rohde1999}), there are many exceptions.  Resolving the finer detail of the apparent upward trend in organismal complexity shows that, under various definitions, complexity has decreased as well as increased across evolutionary time and along branches of the Tree of Life \citep[e.g.,][]{Adamowicz2008,Brinkworth2023,Goldstein2011,MarcotMcShea2007,McShea1991,OMalley2016}.  In addition, complexity is not a reliable guard against adverse outcomes.  In the fossil record, ammonoids developed increasingly elaborate septal sutures, reaching extreme levels of complexity by the Late Cretaceous.  The clade was nevertheless extinguished at the K--Pg boundary, while their close relatives, the much simpler-shelled nautiloids, persisted \citep{Boyajian1992,Saunders2004}.  A similar pattern occurs in crinoids, in which Palaeozoic taxa with highly branched arms and complex calyces were more vulnerable to extinction than simpler forms, and the few stalked crinoids that survive today are relatively conservative \citep{Foote1999,Baumiller2008}.  Graptolites show a comparable pattern: complex, multi-branched diplograptids were lost during the end-Ordovician extinction, while simpler uniserial neograptines survived and diversified in the Silurian \citep{Melchin1991}.  These examples suggest that structural elaboration and specialisation can be advantageous in stable conditions but may render lineages more vulnerable to extinction when environments shift, leaving simpler or more generalist forms as the survivors.  Thus, complexity is far from the infallible answer to survival and replication.  To make progress on the question of why complexity evolves requires understanding the link between complexity and function within a specific context.  The joint investigation of syntactic and semantic complexity aims to achieve and foster further such progress.

\section{Methods}\label{sec:Methods}
\subsection{The general model}\label{sec:Methods_TheGeneralModel}
As a paradigmatic example of a system where syntactic and semantic complexity evolve, we investigate the evolution of model organisms that can be divided into various functional units.  These model organisms are fully described by the capability of their various functional units to perform a variety of functions essential for their survival.  We define the capability of functional unit \(j\) to perform function \(i\) as \(F_i^j\) (\(0\leq F_i^j\leq1\)).  The map from these individual functional capabilities to an organism's capacity for performing a function is given by
\begin{equation*}
    \avep{\mathbf{F}_i}{p}=\left(\sum_jw_i^j\left(F_i^j\right)^p\right)^\frac{1}{p}.
\end{equation*}
\(\mathbf{F}_i=\left(F_i^1,F_i^2,\dots,F_i^N\right)\) is vector of the capabilities of each functional unit \(j\in\{1,2,\dots,N\}\) to perform function \(i\).  \(w_i^j\) are a set of normalised weights (\(\sum_jw_i^j=1\), \(\forall j\)) describing which functional units are ``preferred'' for performing a given task.  The closer \(w_i^j\) is to 1, the more advantageous it is from a fitness perspective for unit \(j\) to evolve specialised adaptations for function \(i\).  When no limb is preferred for performing function \(i\), \(w_i^j=\frac{1}{N}\), \(\forall j\).  \(p\) is a parameter that affects how much large values of \(F_i^j\) are weighted relative to small values.  The extreme values of \(p=\pm\infty\) have the following properties:
\begin{align*}
    \avep{\mathbf{F}_i}{-\infty}&=\min_j\left(F_i^j\right),\\
    \avep{\mathbf{F}_i}{+\infty}&=\max_j\left(F_i^j\right).
\end{align*}
Values of p between these two extremes produce values intermediate to the two above (see Figure~\ref{fig:Figure1}~(c)(iii)).  For example, \(p=1\) results in a standard weighted average:
\begin{equation*}
    \avep{\mathbf{F}_i}{1}=\sum_jw_i^jF_i^j.
\end{equation*}  
This range of behaviour depending on \(p\) can be used to differentiate, for example, between a function that can be performed optimally by a single functional unit (\(p\gg1\)) and a function whose performance requires adaptation across the entire morphology (\(p\sim1\)).

There is a constraint on the maximum functional ability of each functional unit; if a functional unit evolves any adaptation for performing a given function, this uses up part of a finite functional budget, meaning simultaneous optimal adaptation for all functions is not possible.  Note that we assume that adaptation for each function is distinct, meaning there are no correlations between \(F_i^j\).  To implement the constraint on maximum functional ability for a given unit, we require the following inequality to hold:
\begin{equation*}
    \sum_{i}\mathcal{C}_iF_i^j\leq1,\ \forall j.
\end{equation*}
\(\mathcal{C}_i\) is the cost of producing a morphological feature that can perform function \(i\) (\(\mathcal{C}_i>0\)); the greater the value of \(\mathcal{C}_i\), the more limited an organism is in the extent to which a single functional unit can produce adaptations for performing function \(i\).

The fitness of a model organism is related to its ability to perform various functions that are required for its survival.  The organism will be constrained by the function it is least adapted to perform, the limiting factor on its survival.  In this model therefore, the fitness, \(f\), is given by
\begin{equation*}
    f=\avep{\avep{\mathbf{F}_i}{p_i}}{-\infty}\equiv\min_i\left(\avep{\vF_i}{p_i}\right).
\end{equation*}
\(p_i\) is the value of \(p\) corresponding to function \(i\).  Note that a ``softer'' fitness function can easily be incorporated into the model by replacing the value of \(-\infty\) in the fitness function with \(q\), where  \(-\infty<q\ll1\).  Doing this will take into account more than just the minimum functional capability, allowing poor performance at one function to be compensated by good performance at another, to a degree determined by the value of \(q\).  Figure~\ref{fig:Figure1} shows a schematic representation of this model.

\subsection{The general solution}
We wish to find the morphology, given a specific body plan, which maximises fitness.  This is a constrained optimisation problem for which the method of Lagrange multipliers can be used to find a solution.  Formally, the problem is as follows.

\noindent Maximise:
\begin{equation*}
    f=\avep{\avep{\mathbf{F}_i}{p_i}}{-\infty},
\end{equation*}
subject to:
\begin{align*}
    &\sum_i\mathcal{C}_iF_i^j\leq1,\ &&\forall j\in\{1,2,\dots,N\},\\
    &0\leq F_i^j\leq 1,\ &&\forall i\in\{1,2,\dots,n\},\\
    &&&\forall j\in\{1,2,\dots,N\}.
\end{align*}

We need to find the turning points in a Lagrangian containing the fitness function.  To include the constraints, we write down the following Lagrangian:
\begin{multline*}
    \mathcal{L}=\avep{\avep{\mathbf{F}_i}{p_i}}{-\infty}\\
    +\sum_{i=1}^n\sum_{j=1}^N\lambda_{(i-1)N+j}\left(F_i^j-\left(s_{(i-1)N+j}\right)^2\right)\\
    +\sum_{i=1}^n\sum_{j=1}^N\mu_{(i-1)N+j}\left(F_i^j-1+\left(t_{(i-1)N+j}\right)^2\right)\\+\sum_{j=1}^N\nu_j\left(\sum_{i=1}^n\mathcal{C}_iF_i^j-1+\left(u_j\right)^2\right).
\end{multline*}
\(\lambda\), \(\mu\) and \(\nu\) are Lagrange multipliers and \(s\), \(t\) and \(u\) are slack variables, used to impose the inequality constraints.  The first term is the fitness which is to be maximised.  The second term imposes \(F_i^j\geq0\) while the third term imposes \(F_i^j\leq1\).  The final term imposes \(\sum_i\mathcal{C}_iF_i^j\leq1\).  In order to find the optimal fitness, the following equations for all values of \(i\in\{1,2,\dots,n\}\) and \(j\in\{1,2,\dots,N\}\) must be solved simultaneously:
\begin{align*}
    \pdiff{\mathcal{L}}{F_i^j}=0,\\
    \pdiff{\mathcal{L}}{\lambda_{(i-1)N+j}}=0,\\
    \pdiff{\mathcal{L}}{\mu_{(i-1)N+j}}=0,\\
    \pdiff{\mathcal{L}}{\nu_j}=0.
\end{align*}

To begin, we calculate the partial derivative
\begin{align*}
    &\pdiff{\avep{\mathbf{F}_k}{p_k}}{F_i^j}=\pdiff{}{F_i^j}\left(\sum_lw_k^l\left(F_k^l\right)^{p_k}\right)^\frac{1}{p_k},\\
    &=\begin{cases}
        0&\text{if }i\neq k,\\
        w_i^j\left(\frac{F_i^j}{\avep{\mathbf{F}_i}{p_i}}\right)^{p_i-1}&\text{if }i=k.
    \end{cases}
\end{align*}

The partial derivative of the fitness function can be calculated as follows:
\begin{align*}
    &\pdiff{}{F_i^j}\avep{\avep{\mathbf{F}_k}{p_k}}{-\infty}\\
    &=\pdiff{}{F_i^j}\lim_{q\rightarrow-\infty}\left(\frac{1}{n}\sum_{k=1}^n\left(\avep{\mathbf{F}_k}{p_k}\right)^q\right)^\frac{1}{q},\\
    &=\lim_{q\rightarrow-\infty}\left(\frac{\avep{\mathbf{F}_i}{p_i}}{\avep{\avep{\mathbf{F}_k}{p_k}}{q}}\right)^{q-1}\frac{w_i^j}{n}\left(\frac{F_i^j}{\avep{\mathbf{F}_i}{p_i}}\right)^{p_i-1},\\
    &=\delta_{im}w_i^j\left(\frac{F_i^j}{\avep{\mathbf{F}_i}{p_i}}\right)^{p_i-1}=\delta_{im}\pdiff{}{F_i^j}\avep{\mathbf{F}_i}{p_i},
\end{align*}
where \(\delta_{ij}\) is the Kronecker delta and
\begin{equation*}
    \avep{\avep{\mathbf{F}_i}{p_i}}{-\infty}\equiv\avep{\mathbf{F}_m}{p_m}.
\end{equation*}
Thus, we see that
\begin{multline}\label{eq:PartialDerivativeF}
    \pdiff{\mathcal{L}}{F_i^j}=\delta_{im}\pdiff{}{F_i^j}\avep{\mathbf{F}_i}{p_i}\\+\lambda_{(i-1)N+j}+\mu_{(i-1)N+j}+\nu_j\mathcal{C}_i.
\end{multline}
The remaining partial derivatives are
\begin{align}
    &\pdiff{\mathcal{L}}{\lambda_{(i-1)N+j}}=F_i^j-\left(s_{(i-1)N+j}\right)^2,\label{eq:PartialDerivativeLambda}\\
    &\pdiff{\mathcal{L}}{\mu_{(i-1)N+j}}=F_i^j-1+\left(t_{(i-1)N+j}\right)^2,\label{eq:PartialDerivativeMu}\\
    &\pdiff{\mathcal{L}}{\nu_j}=\sum_{i=1}^n\mathcal{C}_iF_i^j-1+(u_j)^2.\label{eq:PartialDerivativeNu}
\end{align}
To find the optimal fitness, the partial derivatives in Equations \eqref{eq:PartialDerivativeF}-\eqref{eq:PartialDerivativeNu} for all values of \(i\in\{1,2,\dots,n\}\) and \(j\in\{1,2,\dots,N\}\) must be set equal to zero and solved simultaneously, repeating this analysis for all combinations of zero (active constraint) and non-zero (inactive constraint) slack variables.  Those combinations which yield feasible solutions are the turning points of the Lagrangian.

\subsection{A simple example}
We investigate a simple example of the model by considering organisms with five functional units, each of which we assume to be a pair of legs.  These leg pairs are required to fulfil three functions: movement, gathering and preparing food and eating.  Gathering and preparing food involves using ``claws'' to break into shells and chop up the contents into edible sized pieces.  Thus, a larger claw is better for performing this function and we assume that a single unit will be able to specialise to perform this task (\(p=\infty\)).  Feeding involves moving adequately prepared food towards the mouth parts for consumption.  We assume that the performance of both feeding and moving will be dependent on adaptation across the whole organism (\(p=1\)).  Thus, leg pair (functional unit) \(i\) is fully described by three values, \(F_1^i=M_i\), \(F_2^i=G_i\) and \(F_3^i=E_i\), representing its moving, gathering and eating ability respectively.  For simplicity, we assume the functional cost of specialisation to each task is equal: \(\mathcal{C}_i=1\), \(\forall i\).  Thus,
\begin{equation*}
    M_i+G_i+E_i=1,\ \forall i.
\end{equation*}
We will also assume that the weights are all equal: \(w_i^j=\frac{1}{N}\), \(\forall i,j\) (\(N=5\) in this example).

To vary the complexity of these organisms, the number of leg pairs constrained to evolve identical adaptations can be altered.  The simplest organism in this model, for example, is constrained such that its leg pairs must all evolve identical adaptations.  Thus, it is defined by two numbers, \(M_1\) and \(G_1\) (with the maximum functional capacity constraint removing a degree of freedom, meaning \(E_1=1-M_1-G_1\)) since the values associated with each leg pair are the same (\(M_i=M_1,\ \forall i\in\{2,3,4,5\}\), etc.).  Higher complexity organisms break this symmetry, allowing various numbers of leg pairs to evolve independent adaptations.  Note that requiring various functional units to evolve the same adaptations is mathematically equivalent to reducing the number of functional units and altering the weights, \(w_i^j\), to reflect the number of units each set of values now represents.

The fitness function for this example of the model is
\begin{equation*}
    f=\avep{\avep{\mathbf{F}_i}{p_i}}{-\infty}=\min\left[\avep{\mathbf{M}}{1},\avep{\mathbf{G}}{\infty},\avep{\mathbf{E}}{1}\right].
\end{equation*}

\subsubsection{Maximally constrained morphology}
In the case of the simplest organism, the fitness function can be simplified since \(\avep{\mathbf{M}}{1}=M_1\), \(\avep{\mathbf{G}}{\infty}=G_1\) and \(\avep{\mathbf{E}}{1}=E_1\):
\begin{equation}\label{eq:FitnessBasic}
f=\min\left[M_1,G_1,E_1\right].
\end{equation}
The fitness landscape described by this reduced fitness function has a single optimum, the point where fitness is maximised, achieved at the morphology described by
\begin{equation*}
    M_1=G_1=E_1.
\end{equation*}
That is, all functions on the right hand side of Equation \eqref{eq:FitnessBasic} are equal.  This is the result of a more general principle which argues that for an organism to be optimally fit, it must have no functional redundancy.  That is, no capacity for performing one function to the extent that any other functional capacity suffers significantly, as this will ultimately limit the fitness of the organism; since the fitness is related to the limiting factor on the organism's survival, exceptional performance at one function is not rewarded by higher fitness if it comes at the cost of capability for performing another function.  Thus, the optimal fitness for a given constrained body plan will feature equal performance of every function (\(\avep{\mathbf{F}_i}{p_i}=\avep{\mathbf{F}_1}{p_1}\), \(\forall i\)).  For the maximally constrained organism, this and the total functional constraint on each limb means
\begin{equation*}
    f=\frac{1}{n}
\end{equation*}
where \(n\) is the number of functions (\(n=3\) in this example).  A cartoon of this morphotype can be seen in Figure~\ref{fig:Figure2}~(a).

\subsubsection{Differentiated front legs}
The complexity of an organism can increase by allowing a constraint to be released.  The next level of complexity up from the most basic organism is one which has two independently evolving functional units (leg pairs).  This complexifying mutation alters the fitness function.  Assuming that our new organism has a front pair of legs which differs from the rear four legs, the fitness function becomes
\begin{equation*}
    f=\min\left[\frac{M_1+4M_2}{5},\max(G_1,G_2),\frac{E_1+4E_2}{5}\right].
\end{equation*}
This complexifying mutation alters the fitness landscape.  The single optimum becomes a saddle point (where \(G_1=G_2\)) and two other turning points appear: a local maximum (where \(G_1<G_2\)) and a global maximum (where \(G_1>G_2\)), such that
\begin{equation*}
    f_{\{G_1>G_2\}}>f_{\{G_1<G_2\}}>f_{\{G_1=G_2\}}.
\end{equation*}
Each fitness is available to a family of morphotypes.  Due to this change to the fitness landscape, an evolutionary path bifurcation occurs, as organisms can evolve towards either the local or global optimum depending on the direction of an initial mutation within this new landscape.  The multi-optima evolutionary landscape in which the organism evolves following such a complexifying mutation is shown in Figure~\ref{fig:Figure2}~(b).  In evolving to the global optimum, the organism's front legs specialise to perform all the gathering, whilst the back legs split their functional capability between moving and eating.  Thus, this complexifying mutation allows division of labour between specialised functional units which is beneficial in this model due to the nature of the gathering function which can be performed optimally by a single functional unit (\(p_2=\infty\)).  Evolving to a local fitness optimum produces a morphology with lower fitness than that of the globally optimal one.  In the extreme case of \(p_1=p_3=1\) and \(p_2=\infty\), there is a manifold of morphologies with exactly equal fitness joining the two local optima seen in Figure~\ref{fig:Figure2}~(b).  For more general parameters, the manifold between the two local optima becomes one of near equal fitness only (see Figure~\ref{fig:Figure2}~(c)).  This family is characterised by specialisation of the four back legs to perform all of the gathering function (opposite to the globally optimal morphotype).  Although this division of labour is advantageous, it is not as advantageous as allowing a single functional unit to perform all the gathering, making morphotypes in this family local rather than global fitness optima.

\subsubsection{Higher complexity}
For a lineage that evolved to the global optimum, further complexifying mutations (releasing further constraints), allowing the back legs to evolve independently, does not give access to higher fitness.  In each case, a single pair of legs specialises to perform all of the gathering function.  Instead of increasing fitness, the change with each complexifying mutation is to the size of the morphospace conferring this optimal fitness (see Figure~\ref{fig:Figure3}~(d)).  With each complexifying mutation, this space increases in size: with two, three, four and five independent leg pairs, the space of optimal morphologies is given by
\begin{align*}
    &N=2:\qquad \mathcal{M}_\mathrm{opt}(M_1,M_2)=0,\\
    &N=3:\qquad \mathcal{M}_\mathrm{opt}(M_1,M_2,M_3)=0,\\
    &N=4:\qquad \mathcal{M}_\mathrm{opt}(M_1,M_2,M_3,M_4)=0,\\
    &N=5:\qquad \mathcal{M}_\mathrm{opt}(M_1,M_2,M_3,M_4,M_5)=0,\\
\end{align*}
where \(\mathcal{M}_\mathrm{opt}\) is a function describing the optimal morphology given constraints.  Clearly, the dimensionality of the optimal morphospace increases with each complexifying mutation.  The dimensionality in each case is given by
\begin{equation*}
    \left(n-1-n_\infty\right)\times\left(N-1\right)
\end{equation*}
where \(n_\infty\) is the number of functions, \(i\), for which \(p_i=\pm\infty\) and \(N\) is the number of independently evolving functional units.  The increase in dimensionality of the optimal morphospace in this example of the model is shown in Figure~\ref{fig:Figure3}~(d).  Note as before that only in the mathematically ideal case of \(p_1=p_3=1\) and \(p_2=\infty\) do the morphotypes contained within the optimal morphospace for each level of complexity (Figure~\ref{fig:Figure3}~(d)) have exactly equal fitness.  For more general parameters the fitnesses of each morphology are nearly equal only.

At each complexity level intermediate to the minimum and maximum, there is a locally optimal fitness state.  This is demonstrated for the first complexifying mutation above.  Similar analysis for three and four independent leg pairs reveals that
\begin{equation*}
    f_{\{G_1<G_2\}}=f_\mathrm{loc,2}<f_\mathrm{loc,3}<f_\mathrm{loc,4},
\end{equation*}
where \(f_{\mathrm{loc,}N}\) is the locally optimal fitness given \(N\) independent leg pairs.  The morphospace in which each fitness can be achieved is given by
\begin{align*}
    &N=2:\qquad \mathcal{M}_\mathrm{loc}(M_1,M_2)=0,\\
    &N=3:\qquad \mathcal{M}_\mathrm{loc}(M_1,M_2,M_3)=0,\\
    &N=4:\qquad \mathcal{M}_\mathrm{loc}(M_1,M_2,M_3,M_4)=0.
\end{align*}
The function \(\mathcal{M}_\mathrm{loc}\) describes the local optimal morphology (such that \(f_\mathrm{loc}<f_\mathrm{opt}\)) given constraints.  There is an advantage to becoming more complex for the lineage stuck in a local fitness optimum as this mutation changes its fitness landscape (the maximum becomes a saddle point), permitting evolution to a new (higher) local fitness optimum or to the global fitness optimum.  Whether the lineage evolves towards the local or global optimum following a complexifying mutation is dependent on the initial specialisation of the newly unconstrained leg pair.  Either this newly independent leg pair will specialise to perform all the gathering function (global optimum) or else the remaining non-independent leg pairs specialise to do this (local optimum).

At every complexifying mutation except the one conferring maximal complexity, there is the possibility of evolving towards a local or global fitness optimum.  Thus, there are multiple evolutionary paths to globally optimal fitness (Figure~\ref{fig:Figure3}~(a)).  The more local fitness optima the path goes through, the higher the syntactic complexity of the organism first evolving a morphology conferring globally optimal fitness.  Semantic complexity increases with syntactic complexity to this point as there is a fitness advantage to higher complexity for an organism in a local fitness optimum.  This also means that the lineage evolves in a mode where complexity is driven to higher values for more rounds of selection, meaning that higher complexity organisms are produced in fewer generations along the evolutionary paths where suboptimal evolutionary branching events have occurred multiple times.

\printbibliography
\end{document}